\documentclass[aps,prl,twocolumn,superscriptaddress,showpacs,amsmath,amssymb,longbibliography]{revtex4-1}
\usepackage{graphicx,bbm}
\usepackage{bm}
\usepackage{color}
\usepackage{hyperref} 

\begin{document}
\title{Distinctive response of many-body localized systems to strong electric field}
\author {Maciej Kozarzewski}
\affiliation{Institute of Physics, University of Silesia, 40-007 Katowice, Poland}
\author{Peter Prelov\v sek}
\affiliation{J. Stefan Institute, SI-1000 Ljubljana, Slovenia }
\affiliation{Arnold Sommerfeld  Center for Theoretical Physics, Ludwig-Maximilians-Universit\" at, D-80333 M\" unchen, Germany }
\author {Marcin Mierzejewski}
\affiliation{Institute of Physics, University of Silesia, 40-007 Katowice, Poland}
\begin{abstract}
We study systems which are close to or within the many-body localized (MBL) regime and are driven by strong electric field.
In the ergodic regime, the disorder extends applicability of the equilibrium linear--response theory to stronger drivings,  whereas
the response  of the MBL systems is very distinctive, revealing currents with damped oscillations.
The oscillation frequency is independent of driving and the damping is not  due to heating but rather due to dephasing. 
The details of damping  depend on the system's history reflecting nonergodicity of the MBL phase,
while the frequency of the oscillations  remains a robust hallmark of localization.  
We show that the distinctive characteristic of the driven MBL phase is also a  logarithmic increase of the energy and the
polarization with time.
 
\end{abstract}
\pacs{71.23.An, 75.10.Pq, 75.10.Jm, 05.60.Gg}



\maketitle
{\it Introduction.--}   The many--body localized  (MBL) systems together with the Anderson insulator \cite{anderson58,fleishman80,kramer93,anderson2}  might represent the only {\em generic} solid--state systems  which do not thermalize \cite{gornyi2005,basko06,oganesyan07,vosk15,altman2015,nonthermal2,rigol,alt,pollmann,huse13} in the thermodynamic limit and may be used to store quantum information \cite{serbyn2014,nonthermal}. 
Among  most characteristic features of the MBL systems are the absense of d.c. transport at any temperature, $T$, \cite{berkelbach10,agarwal15,gopal15}  and the entanglement entropy growing only as a logarithmic function of time \cite{znidaric08,bardarson12,vosk13,serbyn13_1}. 
The  MBL  has recently been identified in optical lattices \cite{schreiber15} by measuring the relaxation dynamics of a particular initial state with charge--density--waves (see also \cite{bordia2015,observ2}). In order to  stimulate further experimental studies, it is essential to specify which hallmarks of the MBL \cite{fabian2015,lev14,sigma2015,monthus10,pal10,dod1,sirker,lauman,pekker} could be detected with well established experimental techniques. 
Several theoretical studies have reported unusual properties of  the optical conductivity, $\sigma(\omega)$, obtained from the linear response (LR) theory   \cite{agarwal15, sigma2015, robin2015,gopal15,barisic10, karahalios09,berkelbach10,herbrych13,barisic16}. 
 The crucial observations concern the low--frequency part of $\sigma(\omega)$ that  goes as $|\omega|^{\alpha}$ with the exponent $1 \le \alpha < 2$ in the MBL state. 

The anomalous linear response of the MBL systems is  important since $\sigma(\omega)$  can be measured via the optical spectroscopy.  
However, it is still unclear when (or whether) the equilibrium LR theory itself is applicable in such systems. It has recently been found that subject to nonzero driving  they go nonlinearly and 
display a highly nonlocal response  at low enough frequencies \cite{Khemani2015}.   Moreover, these systems are expected to posses extensive number of local conserved quantities \cite{integrals1,huse14,integrals2, integrals3,integrals4,vosk13,imbrie2014}
hence, due to these conservation laws, they do not thermalize \cite{nonthermal}, whereas it is common for the LR studies to start from a thermal initial state.  
A first step in clarifying these essential problems is to study the MBL systems driven by a non-zero electric field. Since strong fields drive the  system out of equilibrium,  such studies allow to test not only the linearity of the response but also the consequences of nonthermal initial states.  
 
The evolution of the MBL under strong fields is important also for a general understanding  of driven lattice systems.  Typically, the particle and energy currents  Bloch oscillate with a frequency that is proportional to the field \cite{thermo}, as confirmed for the Falicov--Kimball model \cite{jim2006}, integrable \cite{my1} and nonintegrable \cite{thermo} models of  spinless fermions 
as well as for the Hubbard model \cite{eckstein2011,covaci}. Due to finite d.c. conductivity,  driving causes also the Joule heating that damps  Bloch oscillations.  

In this work, we show that the strong-field response of the MBL systems is very different from the response of standard tight--binding models. Although the particle current undergoes damped oscillations, the frequency is field-independent and the damping is not due to 
heating but mainly due to dephasing. The magnitude of the current as well as the damping depend  on the initial conditions, 
i.e., response within the MBL phase reveals pronounced memory effects, but on the other hand  
the oscillation frequency is independent of the initial (possibly nonthermal)  state and  is a very robust  hallmark of the localization. 
We show furtheron that the oscillations can be well attributed to the local physics within the MBL regime and explained with a local
toy-model. But beyond that our result clearly display  also a logarithmic increase of energy under constant driving which is evidently
a nonlocal effect, having an analogy with a similar slow but steady increase of the entanglement entropy within the MBL phase 
\cite{znidaric08,bardarson12,vosk13,serbyn13_1}.

\begin{figure}
\includegraphics[width=0.49\textwidth]{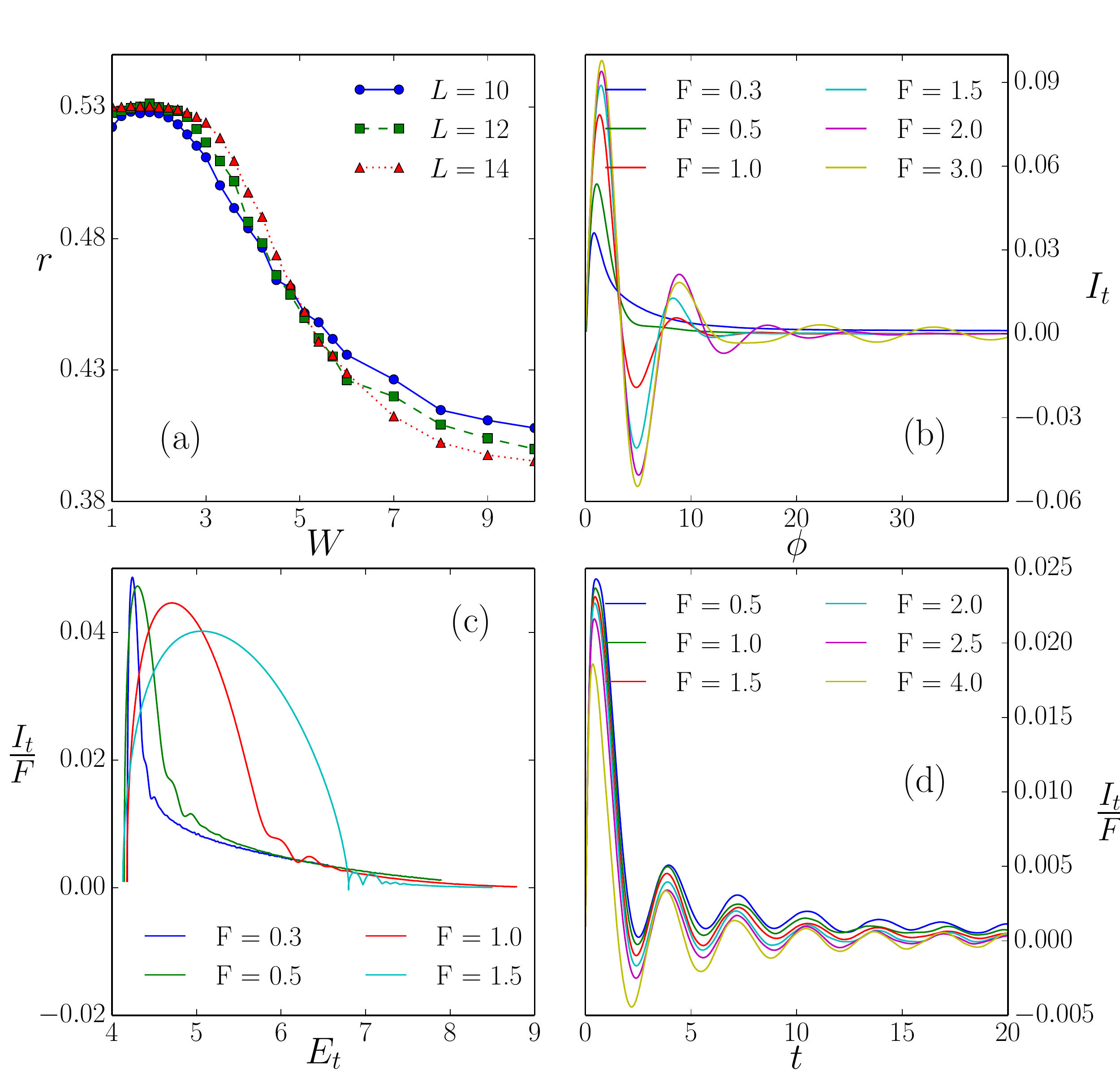}
\caption{ (Color online) (a) The level spacing ratio $r$ vs. disorder $W$; (b)  current $I_t$ vs. $\phi(t)$ for 
weak disorder $W=1$; (c) conductivity $I_t/F$ vs.  instantaneous energy $E_t$ for intermediate $W=3$, and
(d) $I_t/F$ vs. $t$ for $W=6$ within the MBL phase. All results are for the $V=V'=1$. }
\label{fig1}
\end{figure}

 {\it Model--} We study interacting spinless fermions on a one--dimensional lattice with periodic boundary conditions.
 The system is threaded by a time--dependent magnetic flux, $\phi(t)$, which induces the electric  field $F(t)= -\dot{\phi}(t)$. 
The time--dependent Hamiltonian reads,
\begin{eqnarray}
H(t)&=& -t_h \sum_j \left[ {\mathrm e}^{i \phi(t)}\; c^{\dagger}_{j+1}c_j +{\mathrm H.c.} \right] + \sum_j \varepsilon_j \hat{n}_j \nonumber \\
     && + V \sum_j \hat{n}_j \hat{n}_{j+1} +V' \sum_j \hat{n}_j \hat{n}_{j+2},
\label{hi}
\end{eqnarray}
where $\hat{n}_j=c^{\dagger}_{j}c_j $. The hopping integral is taken as the energy unit, $t_h=1$. The potentials  $\varepsilon_j$ have
uncorrelated random values, uniformly distributed in the interval $(-W,W)$.  We have introduced  the nearest--neighbor repulsion, $V$, 
as well as the next-nearest-neighbor repulsion, 
$V'$, so that the ballistic transport is avoided also for $W\rightarrow 0$ \cite{zotos1997}.  
Since most of the previous studies have been carried out for $V'=0$, first we check
 how this interaction affects the MBL transition.  We repeat the analysis  of the energy-level statistics in 
 Refs. \cite{oganesyan07,lev14} and determine the ratio of 
 two consecutive level spacings, $\delta_n$, $\delta_{n+1}$.  In Fig.~\ref{fig1}a we show the average value of the ratio 
 $r =\langle r_n \rangle, r_n= \min\{\delta_n, \delta_{n+1} \} /\max\{\delta_n, \delta_{n+1}\}$ for systems of $L=10, 12, 14$ sites. 
 Upon increasing $W$, we observe a change from $r \simeq 0.53$, consistent with the Wigner-Dyson distribution for ergodic systems,
 to the result of the Poisson-distribution,  $r \simeq 0.39$, which is characteristic for nonergodic 
 (e.g., localized) systems \cite{oganesyan07}. In the thermodynamics limit, the ergodic regime 
 should extend at least upto $W\le 4$ while the MBL should be well visible at least for $W \ge 6$.   
		
{\it Time-evolution --}  We assume that  the field is switched on at $t=0$.
For each set of $\{ \varepsilon_i\}$, the initial state $|\Psi _0 \rangle$ is chosen as a (thermal) microcanonical state \cite{long03} 
with $N=L/2$ fermions and with the energy $E_0$, the latter representing a high-$T$ state. The relation between $E_0$ and $T$
can be then well estimated employing the high--$T$ expansion for the model, Eq.(\ref{hi}), within the canonical ensemble \cite{myrdm},
\begin{equation}
E_0=E_{\infty}- \beta L  \frac{8+V^2+V^{'2}}{16} -\frac{\beta}{4}\sum_i \varepsilon^2_i ,
\label{beff}
\end{equation}
where 
$E_{\infty} \simeq L (V+V')/4$ denotes the  energy at $T\to \infty$ and $\beta=1/T$.  
If not specified otherwise, we choose $\beta=0.2$,  $L=20$ and $V=V'=1$.  The time--evolution of  $|\Psi_t \rangle $ is 
obtained with the help of  the short--iterative Lanczos method \cite{lantime} and the Chebyshev polynomial expansion  
of the time--propagator \cite{chebytime}.  We calculate the energy  $E_t=\langle\langle \Psi_t | H(t) |\Psi_t \rangle \rangle_c$  
and the particle current  $I_t=  \langle \langle \Psi_t | J(t) |\Psi_t \rangle \rangle_c$, where $J=-\frac{d}{d \phi} H(t)/L$ and 
$\langle ... \rangle_c$ represents averaging over disorder configurations.
		
{\it Results --} 
First, we recall  how generic systems respond to constant driving with   $F(t>0)= F$.
In Fig.~\ref{fig1}b we plot $I_t$ as a function of flux $\phi(t)= Ft$  for a  weak disorder. 
Here, $I_t$ vanishes for $t\to\infty$ due to the Joule heating, $\dot{E}_t=FLI_t$ \cite{my1}. For weak $F<1$ one observes
a non--oscillatory decay of $I_t$. On the other hand, for $F\geq 1$  the current  Bloch oscillates as $\sin[\phi(t)]$,  
with a frequency $\omega_B \sim F$. Such  field--dependence of $\omega_B$ is the characteristic feature of  the Bloch oscillations.  
Fig.~\ref{fig1}d  shows seemingly similar behavior for MBL systems in that $I_t$ also undergoes damped oscillations. 
However, the frequency is clearly field--independent, whereas the amplitude is roughly proportional to $F$.  

{\it Ergodic regime --} Before explaining the latter result,  we briefly discuss the case of intermediate disorder, $ 2 \alt W \alt 4$,
which is too weak to cause MBL but strong enough to produce anomalous optical response 
\cite{karahalios09,agarwal15, gopal15, robin2015,barisic16}.  
Such systems are ergodic and relax towards the thermal state, hence the only concern  related to the applicability 
of the  equilibrium $LR$ theory is whether the response is indeed linear in $F$.  Strictly speaking, even a slow Joule heating is 
a nonlinear effect which, however,  can be easily accounted for within a simple extension of the LR theory 
\cite{my1}. A convenient way to filter out the heating effect is to plot the observables as a function 
of the instantaneous energy $E_t$ (see Fig.\ref{fig1}c).  For modest driving the system then undergoes a quasi-thermal 
evolution, i.e.,  the time-dependent expectation values of all local operators  are expected to be determined solely by  
$E_t$ \cite{myrdm}.   Fig.~\ref{fig1}c shows that in the long--time regime,  
the effective conductivity $I_t/F$ for intermediate $W=3$ is indeed uniquely determined by $E_t$ and 
roughly $F$--independent.  Comparing further results  for 
moderate driving  ($F \sim 1$) one finds  non--oscillatory linear response for intermediate $W=3$  (Fig.~\ref{fig1}c) 
and  very clear Bloch oscillation for weak disorder $W=1$  (Fig.~\ref{fig1}b). So our  main conclusion 
for the ergodic phase is  that  the LR theory is applicable to much larger fields  in more 
disordered systems.
 
\begin{figure}
\includegraphics[width=0.49\textwidth]{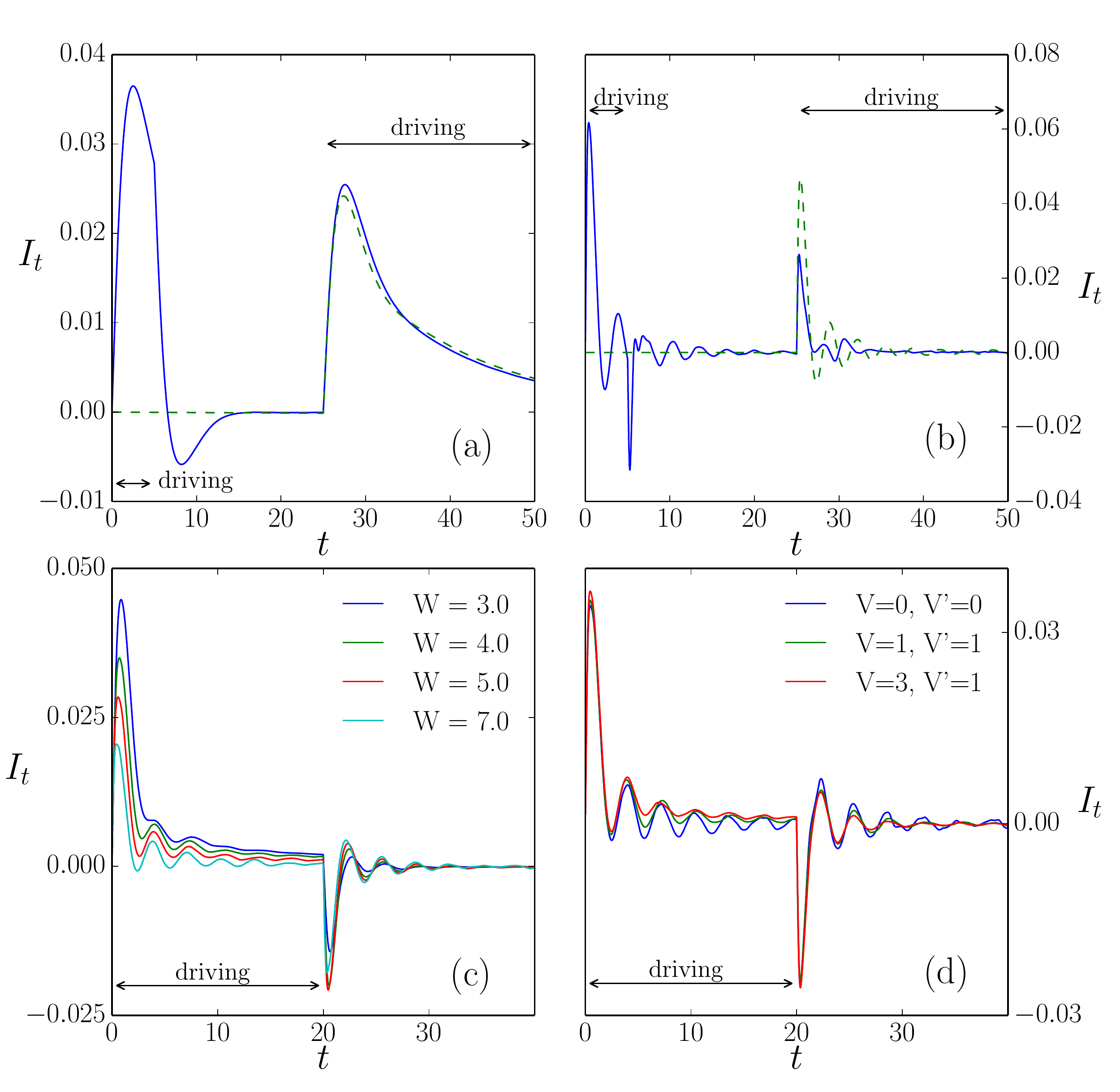}
\caption{ (Color online) Current $I_t$ vs. $t$ for various parameters and driving protocols. Horizontal arrows mark time windows where $F\ne 0$:
(a) weak $W=1$ and $F=0.3$, and (b) large $W=6$ and $F=3$ in marked time slots, respectively 
(Dashed lines in (a) and (b) show response to analogous second pulse 
but for systems which up to $t=t_0=25$ are in thermal states); 
(c) fixed driving $F=1$ and various $W$; (d) fixed $W=6$  and $F=1.5$ but different $V,V'$. }
\label{fig2}
\end{figure}
 
{\it MBL regime - memory effects--} A particularly interesting aspect of  MBL are the memory effects. 
Since MBL systems do not thermalize, their response may depend on the history of the system, 
in particular, whether it was previously driven out of equilibrium. It order to study this effect, we turn off the driving for a 
time interval such that the transient particle current relaxes $I_t \sim 0$, and  then turn on the field  again at $t=t_0$.  
Evolution of $I_t$ under such specific driving is shown by continuous lines
in  Figs.~\ref{fig2}a and b for the ergodic and MBL regimes, respectively.  
For comparison we present also the effect of the second pulse provided that the pulse excites the system 
within the thermal (microcanonical) state with the same energy (see dashed curves). It clearly follows, 
that in contrast to the ergodic case,  the MBL regime has pronounced memory effect, i.e. the response strongly 
depends  on the initial conditions.
    
{\it Current oscillations--} Furtheron we return to oscillations of $I_t$ close to or within the MBL regime.  
Results presented in  Figs.~\ref{fig2}c,d reveal that
the frequency $\omega_0 \simeq 2$  is a very robust property of  our disordered model system.  
In particular, $\omega_0$ is independent of  $W$ as well as of interactions $V,V'$, as seen in  
Figs.~\ref{fig2}c,d.  
We note, that weak signatures of damped 
oscillations are visible even in the ergodic phase for $W \sim 3 $ (see Fig. \ref{fig2}c).

 \begin{figure}
\includegraphics[width=0.49\textwidth]{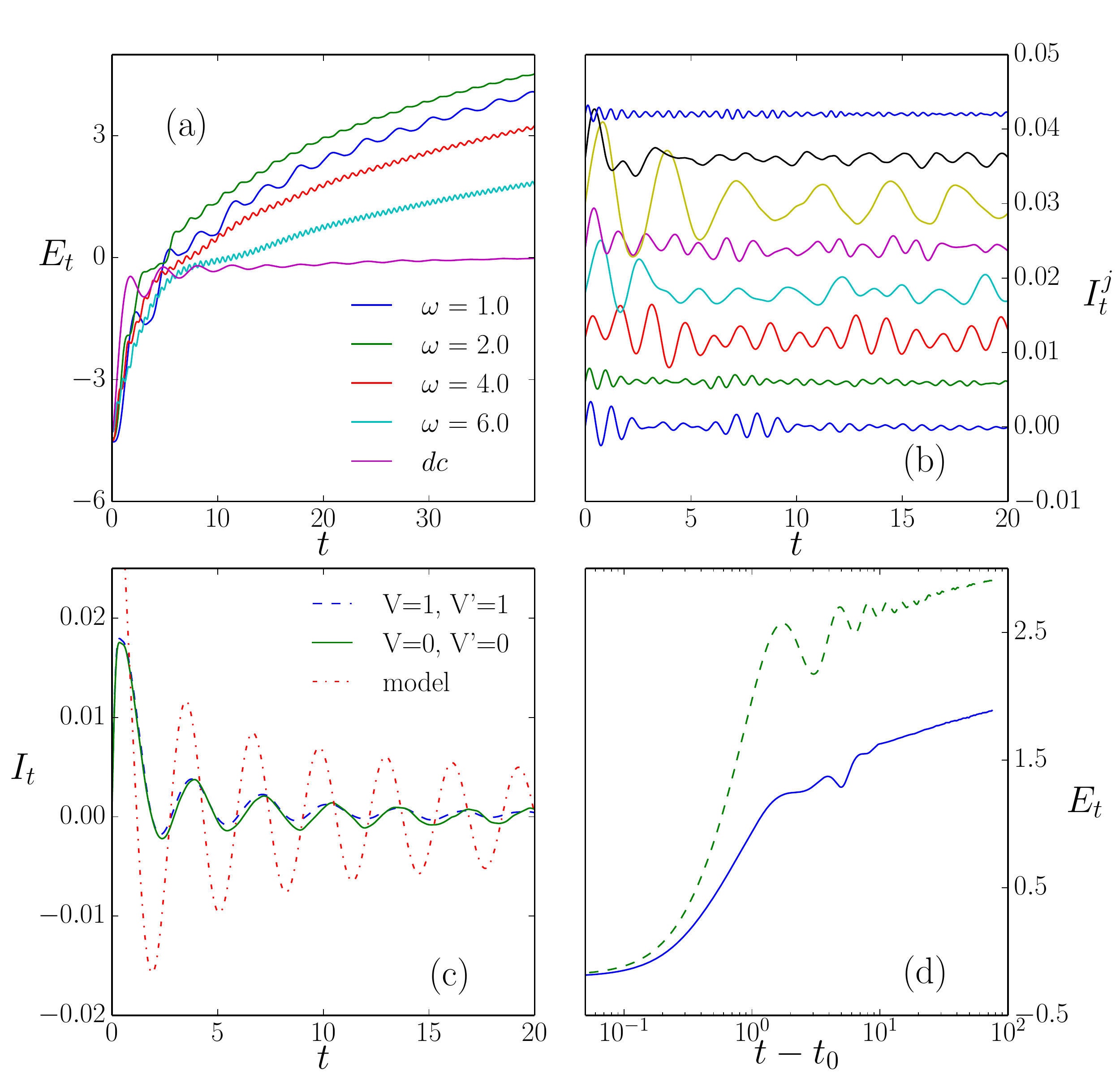}
\caption{ (Color online) (a) Energy $E_t$ vs. $t$ for a.c. driving $F(t>0)=3\sin(\omega t)$ including d.c.  case
$F=3$, all for $W=6$; (b)  local currents $I_j^t$ on consecutive bonds  (shifted vertically for clarity) 
for d.c. $F = 3$ within the MBL regime,   $W=6$;  (c) numerically obtained $I_t$ for  $F=1$, $W=8$ within the interacting and  noninteracting models, respectively, compared with result from 
the toy-model. (d) Long-time variation of $E_t$ for the same cases as in Fig. \ref{fig2}b.  
All results except c) are for $V=V'=1$.}
\label{fig3}
\end{figure}

In order to go deeper into the physics behind these oscillation, we study also a.c. driving with  
$F(t>0)=F \sin(\omega t)$. Other type of periodic driving has   been studied in Refs.~\cite{periodic1,periodic2}. 
Fig.~\ref{fig3}a shows that  the strongest absorption of energy is exactly for $\omega=\omega_0$. 
We find  for such driving that $E_t$  increases and eventually approaches the $T=\infty$ value
($E_{\infty} \approx 10$ for parameters in Fig. \ref{fig3}a), whereas for the d.c. driving the energy 
apparently saturates at much lower values (see more detailed analysis furtheron).  
This last observation can be reconciled with the LR result for the MBL regime,
$\sigma_{dc}\sim 0$,  which 
remains  qualitatively  valid  even for strong fields $F>1$, as shown in Fig.~\ref{fig3}a. 

A clearer picture of the oscillations arises from comparing the currents flowing on individual bonds, 
 \begin{equation}    
 I^j_t =  \langle \Psi_t | ( i   \mathrm{e}^{i \phi(t)}  c^{\dagger}_{j+1}c_j  + \mathrm{c.c.}) |\Psi_t \rangle ,  \label{iti}
\end{equation}
 so that   $I_t=\sum_j  \langle I^j_t \rangle_c /L$.
 In the ergodic phase, the currents on the neighboring bonds, $I^j_t$ and  $I^{j+1}_t$, are quite correlated with each other (not shown). 
  However in the MBL regime, $I^j_t$ and  $I^{j+1}_t$ oscillate with very different frequencies and magnitudes, as shown in Fig. \ref{fig3}b. 
 In contrast to $I_t$, the damping of currents on individual  bonds is hardly visible. The latter result clearly indicates that damping of $I_t$
 is actually due to destructive interference of various $I^j_t$.
 
{\it Toy model.} Within the MBL phase, the currents on neighbouring bonds appear to be independent of each other. 
This suggests that results for decoupled two--site clusters should capture the essential physics.   
Therefore, we briefly discuss a toy-model on two sites with the following  Hamiltonian  and the current operator
\begin{equation}
H_2(t)=\left(
\begin{array}{cc}
\epsilon+\frac{F(t)}{2} & 1 \\1 & -\epsilon-\frac{F(t)}{2} \\
\end{array}
\right),\quad J_2=\left(\begin{array}{cc}0 & i\\ -i &0 \\
\end{array}
\right).
\end{equation}  
The distribution of  random  $\epsilon$, depending on $W$, can also incorporate  the many--body interaction 
between neighboring clusters. Here, we assume only that the probability density is even $f_{\epsilon} =f_{- \epsilon} $.
An arbitrary initial state can be written as   
\begin{equation}
\rho(0)=x |\phi_-\rangle \langle \phi_- |+(1-x) |\phi_+\rangle \langle \phi_+| +\left(\alpha |\phi_-\rangle \langle \phi_+|+\mathrm{H.c.}\right), \label{ini}
\end{equation}  
where $(x-1/2)^2+|\alpha|^2 \le 1/4$, while $| \phi_{\pm} \rangle $ are eigenstates of $H_2(0)$ with energies $\pm \sqrt{1+\epsilon^2}$. 
In general, $x$ and $\alpha$ may depend on $\epsilon$, e.g., for the thermal state one obtains $\alpha=0$ and  
$x_{\epsilon}=1/2+\tanh(\beta \sqrt{1+\epsilon^2})/2$.
We assume that  $x_{\epsilon}=x_{-\epsilon}$ and $\alpha_{\epsilon}=\alpha_{-\epsilon}$.
Then, straightforward calculations show that driving $F(t)=F \theta(t)$ induces the current (given here only up to the linear term in $F$),
\begin{eqnarray}
I_t &= & \langle \mathrm{Tr}[\rho(t)J_2] \rangle_c=\langle  I^0_t + I^F_t  \rangle_c+O(F^2), \label{i2} \\
I^0_t &=&  -2 \Im [ \alpha \exp( i\;2 \sqrt{1+\epsilon^2}\;  t)]  , \label{i0} \\
I^F_t &= &   \left(x-\frac{1}{2}\right) \frac{\sin(2\sqrt{1+\epsilon^2} \; t)}{1+\epsilon^2} F. \label{i1} 
\end{eqnarray}
$I^0_t$  is independent of driving and arises solely due to non--steady  initial conditions ($\alpha \ne 0$), 
whereas $I^F_t$ describes the LR response. Eq.~(\ref{i1}) explains,  at least qualitatively, why  the largest 
amplitudes of $I^j_t$ shown in Fig.~\ref{fig3}b  are oscillating with the smallest frequencies. 
In the long-time regime, the disorder--averaged $\langle I^F_t\rangle_c$ can be obtained analytically for 
arbitrary  $x_{\epsilon}$ and $f_{\epsilon}$. The asymptotic  form is then
\begin{equation}
\langle I^F_t\rangle_c = \sqrt{\frac{\pi}{t}} f_{\epsilon=0} \left(x_{\epsilon=0}- \frac{1}{2}\right)  F \sin \left(2t+\frac{\pi}{4} \right) .
\label{toy}
\end{equation}
The average current oscillates with the smallest possible frequency, $\omega_0=2$,  and decays slowly in time as $1/\sqrt{t}$ 
due to destructive interference of oscillations with different frequencies (as seen in Fig.~\ref{fig3}b).   
Fig.~\ref{fig3}c shows numerical results for the original Hamiltonian (\ref{hi}) compared with $\langle I^F_t \rangle_c$  
obtained from Eq.~(\ref{toy}) for $f_0=1/W$ and $x_0=1/2+\tanh(\beta)/2$. The toy-model is too simple 
to describe details of the damping which appears from Fig.~\ref{fig2}d to be mostly determined by the many--body interactions. 
However, the toy-model correctly reproduces the specific frequency $\omega_0$ of these oscillations. 
Most importantly, it  explains also why the same frequency is obtained for various types of disorder ($f_{\epsilon} $) 
and various (also nonthermal)   initial conditions ($\alpha_{\epsilon},x_{\epsilon}$). 

{\it Logarithmic increase of energy and polarization--}  Within the fully localized regime one expects 
that the driving with constant field $F$ would finally lead to the saturation (or oscillation) of various quantities.
This is indeed the case for the noninteracting Anderson model. A more detailed analysis of the MBL results,
however, reveals a deviation at long times. In Fig.~\ref{fig3}d we present $E_t$ in the long--time window for exactly the same cases
as in Fig. \ref{fig2}b, i.e. for thermal and non--thermal initial states.  In both cases, one observes a slow steady growth of  $E_t$ even for $t\gg 1$, 
which is consistent with the logarithmic dependence, $\Delta E_t \propto \mathrm{log}(t)$.  Since the increase of the  energy is exactly related to the current 
\cite{my1} as $\dot{E}_t=L I_t F(t)$, upon constant $F(t)=F$ one can directly test the plausible relation
$\Delta E_t= L \Delta P_t F$, where $P_t$ is the polarization of the system. Hence, the observed variation
implies as well the dependence $\Delta P_{t\gg 1}  \propto  \mathrm{log}(t)$.  The growth without an upper
bound opposes the plausible picture that MBL insulator is a dielectric with a finite polarizabiltity $\chi=P/F$,
still it is consistent with recent findings \cite{agarwal15,robin2015,barisic16} 
that the low-frequency dynamical conductivity beyond the transition to the MBL phase behaves as $\sigma(\omega) \propto  \omega^\alpha$
with $\alpha \sim 1$. Namely, at least within the LR theory we would get $\chi \propto \int d \omega \sigma(\omega)/
\omega^2$ which for $\alpha=1$ diverges logarithmically.  Similarly to the particle current, also the polarization shows strong memory effects, 
however  the logarithmic character of its growth seems to be independent of the initial conditions, as shown in Fig. ~\ref{fig3}d.  
 
 {\it Conclusions--}  
 We have identified distinctive properties of the MBL systems  driven by non--zero (strong) electric field.
They can be classified according to locality and linearity of the underlying physics:

\noindent (a) The oscillations of the particle current with a  field-independent frequency  are very pronounced already within the crossover to the MBL phase 
and even more deeper within it. Our toy--model confirms that they are the consequence of very local (two-site) physics, while
decay of oscillations in the nonergodic (MBL phase) is due to the dephasing. Such oscillations might be observed also in
recent experiments \cite{bordia2015}.

\noindent (b) We find within the MBL phase very pronounced nontrivial memory effects  which typically reveal physics beyond the equilibrium LR.

\noindent (c) In contrast to above phenomena, the observed logarithmic increase of energy and polarization with time
under constant driving clearly goes beyond the local physics. While it is consistent with the LR anomalous conductivity 
$\sigma(\omega)\propto |\omega|$  at the MBL transition (or crossover) \cite{agarwal15,gopal15,robin2015,barisic16}, 
our results show that it remains valid also well beyond LR. It appears to go along with similar anomalies as e.g. the
entanglement entropy also growing $S \propto \mathrm{log}(t)$ \cite{znidaric08,bardarson12,vosk13},
which are specific hallmarks of the MBL phase, but so far only partially understood.

\acknowledgements
M.M. acknowledges support from the DEC-2013/09/B/ST3/01659 project of the Polish National Science Center.  
P.P. acknowledges support by the program P1-0044 of the Slovenian Research Agency and by the Alexander von 
Humboldt Foundation.

\bibliography{ref_mbl.bib} 

\end{document}